%% file: 6D_grav_lhc.tex
\documentclass[onecolumn,prd,superscriptaddress,showpacs,floatfix, 
preprintnumbers,nofootinbib,showkeys]{revtex4-1}  

\usepackage{amsmath,amsfonts}
\usepackage{epsfig,epsf}
\usepackage{mathrsfs}
\usepackage{graphicx}
\usepackage{latexsym}
\usepackage{amsmath}
\usepackage{amssymb}
\usepackage{textcomp}
\usepackage{amsbsy}
\def\beq{\begin{equation}}
\def\eeq{\end{equation}}
\def\barr{\begin{array}}
\def\earr{\end{array}}
\def\dis{\displaystyle}
\def\lapp{\mathrel{\rlap{\raise.5ex\hbox{$<$}}
                    {\lower.5ex\hbox{$\sim$}}}}
\def\gapp{\mathrel{\rlap{\raise.5ex\hbox{$>$}}
                    {\lower.5ex\hbox{$\sim$}}}}
\def\td{\widetilde \Delta}
\def\tev{\, {\rm TeV}}

\usepackage[dvips]{color}
\usepackage[normalem]{ulem}
\definecolor{darkgreen}{cmyk}{1,0,1,0.4}

\definecolor{darkred}{cmyk}{0,1,1,0.4}

\begin{document}

\title{\large \bf Graviton modes in multiply warped geometry}

\author{Mathew Thomas Arun}
\email{thomas.mathewarun@gmail.com}

\author{Debajyoti Choudhury}
\email{debajyoti.choudhury@gmail.com}
\affiliation{Department of Physics and Astrophysics, University of Delhi, 
Delhi 110007, India.}

\author{Ashmita Das}
\email{ashmita.phy@gmail.com}

\author{Soumitra SenGupta}
\email{soumitraiacs@gmail.com}
\affiliation{Department of Theoretical Physics, Indian Association for the
Cultivation of Sciences, 2A\&B R.S.C. Mullick Road, Kolkata 700 032, India }

\date{\today}

\begin{abstract}
The negative results in the search for Kaluza-Klein graviton modes at
the LHC, when confronted with the discovery of the Higgs, has been
construed to have severely limited the efficacy of the Randall-Sundrum
model as an explanation of the hierarchy problem. We show, though,
that the presence of multiple warping offers a natural resolution 
of this conundrum through modifications in both the graviton spectrum 
and their couplings to the Standard Model fields.

\end{abstract}

\maketitle

\section{Introduction}
Despite the spectacular success of the Standard Model (SM) of
elementary particles, the search for new physics beyond the SM
continues. One of the primary motivations for this is to resolve the
well-known gauge hierarchy/naturalness problem in connection with the
fine tuning of the higgs mass against large radiative corrections.
Among several proposals to address this problem, models with extra
spatial dimensions draw special attention. In this context, the warped
geometry model proposed by Randall and Sundrum (RS) 
\cite{RS} turned out to
be particularly successful for $(i)$ it
resolves the gauge hierarchy problem without bringing in any other
intermediate scale in the theory in contrast to the large extra
dimensional models; $(ii)$ the modulus of the extra dimensional model
can be stabilized to a desired value by the Goldberger-Wise mechanism
\cite{GW1}, and $(iii)$ a similar warped solution can be obtained from
a more fundamental theory like string theory where extra dimensions
appear naturally\cite{Green}.  As a result, several search strategies
at the LHC were designed specifically \cite{ATLAS1, Aad:2012cy,ATLAS:2013jma,CMS1} to detect the
indirect/direct signatures of these warped extra dimensions
e.g. through the dileptonic decays of 
Kaluza-Klein (KK) excitations of the graviton which
appear in these models at the TeV scale. 

The original RS model was defined as a slice of AdS$_5$ space with a
$S^1/Z_2$ orbifolding and a pair of three-branes located at the
orbifold fixed points, viz. $y = 0, \pi$ (with the SM fields being
localized on the last mentioned).  The parameters characterizing the
theory are the 5-dimensional fundamental (gravitational) scale $M_5$
and the bulk cosmological constant $\Lambda_5$. The solution to the
Einstein's equations, on demanding a $(1+3)$--dimensional Lorentz
symmetry, then leads to a warp-factor in the metric of the form
$\exp(-k_5 \, r_c \, y)$ where $r_c$ is the compactification radius
and $k_5 = \sqrt{-\Lambda_5 / 24 \, M_5^3}$.  Clearly, the
applicability of the semiclassical treatment (as opposed to a full
quantum gravity calculation) requires that the bulk curvature $k_5$ be
substantially smaller than $M_5$.  An analogous string theoretic
argument~\cite{DRgr} relating the D3 brane tension to the string scale
(related, in turn, to $M_5$ through Yang-Mills gauge couplings) demand
the same, leading to $k_5/M_5 \lapp 0.1$.  On the other hand, too
small a value for this ratio would, typically, necessitate a
considerable hierarchy between $r_c^{-1}$ and $M_5$, thereby taking
away from the merits of the scenario. Thus, it is normally accepted
that one should consider only $0.01\leq k_5/M_5 \leq 0.1$.  Indeed,
this constraint plays a crucial role in most of the phenomenological
studies of this scenario, and certainly for the aforementioned results
reported by the ATLAS and the CMS groups.  Throughout our analysis we
shall impose an analogous condition on the bulk curvature as an
important restriction to ensure the applicability of our semiclassical
calculations.

In the context of the original RS model, the large
exponential warping is held responsible for the apparent lightness of
the Higgs vacuum expectation value $v$ (and its mass), as
perceived on our brane, related as it is to some naturally high scale
$\widetilde v \sim {\cal O}(M_5)$, applicable at the other brane,
through the relation
\begin{equation}
  v  = \widetilde v \, e^{- \pi \, k_5 \, r_c } \ .
          \label{5D_higgs}
\end{equation}
Here $\widetilde v$ is determined by the natural scale of higher dimensional model $\sim $ five dimensional Planck scale $M_5$ and $k_5 \, r_c \approx 12$ would explain the hierarchy with $r_c$
being stabilized to this value by some mechanism~\cite{GW1}. The
compactification leads to a nontrivial KK tower of gravitons with the
levels being given by
\begin{equation}
m_n \, = \, x_n \, k_5 \, e^{-\pi \, k_5 \, r_c}
      \label{5D_spectrum}
\end{equation}
where $x_n$'s are the roots of the Bessel function of order one. With
only the lowest (massless) graviton wavefunction being localized away
from our brane, its coupling to the SM fields is small, viz. ${\cal
  O}(M_5^{-1})$. As the couplings of the others to
the SM fields suffer no such suppression, they
are, presumably, accessible to collider searches. The ATLAS
collaboration~\cite{Aad:2012cy}, though, has reported negative results
ruling out a level--$1$ KK graviton in the mass range below 1.03
(2.23) TeV, with the exact lower bound depending on the value chosen
for $k_5 / M_5$.

This result immediately brings forth a potential problem for the model, 
for eqns.(\ref{5D_higgs}\&\ref{5D_spectrum}) together demand
that 
\begin{equation}
   \frac{m_1}{m_H} \sim \frac{m_1}{v} =  x_1 \, \frac{k_5}{\widetilde v} \ =  x_1 \, \frac{k_5}{M_5}\frac{M_5}{\widetilde v} \
\end{equation}
Since $k_5 / M_5 \lapp 0.1$, it is immediately apparent that, unless
$\widetilde v$ is at least two orders of magnitude smaller than $M_5$,
a 126 GeV Higgs~\cite{ATLASHiggs, CMSHiggs} would cry out for a KK
graviton below a TeV. Indeed, this argument has been inverted in the
literature\cite{Das:2013lqa} to argue for a much lower cutoff (in
other words $\widetilde v$) in the theory. In other words, some new
physics would need to appear at least two orders of magnitude below
the fundamental scale $M_5$, which, in the RS scenario is very close
to the four-dimensional Planck scale itself.

Let us remind ourselves of the
  nature of cutoffs in the effective four-dimensional theory,
  considered as a theory of the the SM fields augmented by the RS
  gravitons. While the SM is operative below the scale of the first KK
  graviton, the {\em new} four-dimensional theory is operative all the
  way up to the compactification scale $\sim r_{c}^{-1}$ when each of the KK graviton is expected to  take part in the amplitude estimation as the 
beam energy is increased. Beyond the energy  $\sim r_{c}^{-1}$, we indeed encounter new physics by probing into
the extra dimension where the theory can no longer be defined as an
effective theory in four dimensions defined by standard model and KK gravitons.  

It is important to realize, at this stage, that part of the
aforementioned problem lies in the very restrictive nature of the RS
model as it is impossible to lower $r_c^{-1}$ by two orders without
disturbing the value of the warped factor significantly.  This, in
turn, would introduce a little hierarchy necessitating a fine tuning of 2-3 orders so that the
Higgs mass may be kept $\sim 125$ GeV.  This feature would worsen
further if a graviton KK mode continues to elude us in the forthcoming runs of the 
LHC, as well as in future collider experimnts.

On the other hand, within the context of a
  generalization  of the RS model with additional warped
extra dimensions, a lower cutoff appears {\em naturally}, 
in the form of a larger compactification radius. 
In other words, the problem is circumvented without the need 
for any additional (small) fine tuning.
Indeed, once we admit more than four dimensions, there is no
particular reason to restrict the number to five, especially with
constructs such as string theoretic models arguing in favour of many
more.  Such variants of the RS model have been proposed earlier
\cite{Shaposhnikov,Nelson,Cohen,Giovannini,dcssg} with these,
typically, considering several independent $S^1/Z_2$ orbifolded
dimensions along with $M^{(1,3)}$.  For example, codimension-2 brane
models \cite{Kanti} have been invoked to address aspects like Hubble
expansion and inflation \cite{Terning,Shiu,ChenF}, Casimir densities \cite{Saharian:2006jv,Saharian2}, little RS hierarchy \cite{McDonald:2008ss},
gravity and matter field localizations \cite{Midodashvili:2004nt,Rizzo}, fermion mass generations \cite{Grossman,Perez},
moduli stabilization \cite{Tanaka}, etc. \\
 We begin our study, with a
brief discussion of the basic features of warped geometry model in
6-dimension with two succesive $S_1/Z_2$ orbifoldings.

\section{Multiply warped brane world model in 6D}    
Consider a doubly warped compactified six-dimensional space-time 
with successive $Z_2$ orbifolding in each of the extra dimensions, viz. 
$M^{1,5}\rightarrow[M^{1,3}\times S^1/Z_2]\times S^1/Z_2$. Demanding 
four-dimensional ($x^\mu)$ Lorentz symmetry within the set up, requires the 
line element to be given by~\cite{dcssg}
\begin{equation}
\label{metric}
ds^2_6= b^2(z)[a^2(y)\eta_{\mu\nu}dx^{\mu}dx^{\nu}+R_y^2dy^2]+r_z^2dz^2 \ ,
\end{equation}  
where the compact directions are represented by the angular
coordinates $y, z \in [0,\pi]$ with $R_y$ and $r_z$ being the
corresponding moduli. Just as in the RS case, nontrivial warp factors
$a(y)$ and $b(z)$, when accompanied by the orbifolding necessitates
the presence of localized energy densities at the orbifold fixed
points, and in the present case, these appear in the form of tensions
associated with the four end-of-the-world 4-branes. 

The total bulk-brane action for the six dimensional space time is, thus,
\begin{equation}
\barr{rcl}
{\cal{S}}&=& \dis {\cal{S}}_6+{\cal{S}}_5 \\[1.5ex]
{\cal{S}}_6&=& \dis \int d^4x \, dy \, dz \sqrt{-g_6} \, 
   (M_{6}^4R_6-\Lambda)\\[1.5ex]
{\cal{S}}_5&=& \dis \int d^4x \, dy \, dz \sqrt{-g_5}\, 
      [V_1(z) \, \delta(y)+V_2(z) \, \delta(y-\pi)]\\
&+& \dis \int d^4x \, dy , dz\sqrt{-\tilde g_5} \, 
     [V_3(y) \, \delta(z)+V_4(y) \, \delta(z-\pi)] \ ,
\earr
\end{equation}
where $\Lambda$ is the (six dimensional) bulk cosmological constant and $M_6$ is the
natural scale (quantum gravity scale) in six dimensions. The
five-dimensional metrics in ${\cal S}_5$ are those induced on the
appropriate 4-branes, which accord a rectangular box shape to the
space.  Furthermore, the SM (and other) fields may be localized on
additional 3-branes located at the four corners of the box, viz.
\[
{\cal{S}}_4 = \sum_{y_i, z_i = 0, \pi} \, 
\int d^4x \, dy \, dz \, 
\sqrt{-g_{4}} \, {\cal L}_i \, \delta(y- y_i) \, \delta(z- z_i) \ .
\]
These terms, however, are not germane to the discussions of this
paper, and we shall not discuss ${\cal{S}}_4$ any further.

For a negative bulk cosmological constant $\Lambda$, the solutions for
the 6-dimensional Einstein field equations are given by~\cite{dcssg}
\begin{equation}
\barr{rclcrcl}
a(y)& = & e^{-c|y|} &\hspace{2cm}& c & = & \dis \frac{R_y k}{r_z\cosh{k\pi}}
\\[1ex]
b(z)& = & \dis \frac{\cosh{(kz)}}{\cosh{(k\pi)}} &\hspace{2cm}& 
k& = & \dis r_z\sqrt{\frac{-\Lambda}{10 M_6^4}} \equiv r_z \,k' \ .
\earr
                \label{RS6_eqns}
\end{equation}
The Israel junction conditions specify the brane tensions. The
smoothness of the warp factor at $z = 0$ implies $V_3(y)$ be
vanishing, while the fixed point at $z = \pi$ necessitates a negative
tension, viz.
\begin{equation}
\label{zten}
V_3(y)=0,\hspace{1cm} V_4(y)=\frac{- 8M^4k}{r_z} \,\tanh{(k\pi)} \ .
\end{equation}
With the warping in the $y$-direction being similar 
to that in the 5D RS model, the two 4-branes sitting at 
$y = 0$ and $y = \pi$ have equal and
opposite energy densities. However, the $z$-warping dictates 
that, rather than being constants, these energy densities must 
be $z$-dependent, viz.
\begin{equation}
\label{yten}
V_1(z)=-V_2(z)=8M^2\sqrt{\frac{-\Lambda}{10}} \, {\rm sech}(kz) \ .
\end{equation}
Such a $z$-dependence can arise from a scalar field distribution confined on the brane. For a
detailed discussion on this we refer our reader to section III of \cite{dcssg}.
The (derived) 4-dimensional Planck scale can be related 
to the fundamental scale $M$ through 
\begin{equation}
   M_P^2 \sim \frac{M_6^4 \, r_z \, R_y}{2 \, c \, k}  \,
              \left(1 - e^{-2 \, c \, \pi} \right) \;
              \left[ \frac{\tanh k \,\pi}{\cosh^2 k \, \pi} 
                   + \frac{\tanh^3 k \,\pi}{3}  \right] \ .
    \label{Planck_mass}
\end{equation}
If there exists no other brane with an energy scale lower than ours,
we must identify the SM brane with the one at $y = \pi, z = 0$. This
immediately gives the required hierarchy factor ({\em i.e.} the mass
rescaling due to warping) to be
\begin{equation}
\label{hierarchy}
w =  \frac{e^{-c \pi}}{\cosh{k\pi}} \ .
\end{equation}
For the large hierarchy that we need to explain, this equation,
alongwith the relation between $c$ and $k$ (eqn.\ref{RS6_eqns})
demands that, unless there is a very large hierarchy between the
moduli, the warping is substantial in only one of the two directions,
and rather subdominant in the other.  In other words, we can have
either ($i$) a large ($\sim 10$) value for $k$ accompanied by an
infinitesimally small $c$ or ($ii$) a large ($\sim 10$) value for $c$
with a moderately small ($\lapp 0.3$) $k$.  The issue of moduli
stabilization in such multiple moduli scenario is yet to be addressed.
However, in view of the essential similarity of the warp factors to
the RS case, we believe that an analogue of the Goldberger-Wise
stabilization mechanism \cite{GW1}, using either a bulk
six-dimensional scalar field, or a combination of 4-brane localized
scalars would fit the bill. This is currently under investigation.

In summary, we are dealing with a brane world which is doubly warped,
with the warping being large along one direction and small in the
other.  The very structure of the theory typically requires a small
hierarchy between the two moduli, both of which remain comparable to
the fundamental length scale in the theory.  The stability issues in
such models have been studied along with the effects of bulk gauge
field or higher form anti-symmetric tensor
field\cite{Sdas,Alberghi,GW2,Behrndt}.

Apart from the gauge hierarchy problem, such a model can offer a
possible resolution of the observed fermion mass hierarchy \cite{Rssg}.
 Furthermore, we can achieve a consistent description
of a bulk Higgs and gauge fields with spontaneous symmetry breaking in
the bulk, along with proper $W$ and $Z$ boson masses on the visible
brane \cite{hsd}.  Given these successes of the model, it is
interesting to consider the graviton sector of the theory and, in
particular, to investigate whether it is consistent with the LHC
bounds.

\section{The graviton KK modes}

To obtain the KK modes, one needs to consider the fluctuations of the metric,
\begin{equation}
g_{MN} = \bar g_{MN} + \Delta_{MN} 
\end{equation}
where $\bar g_{MN}$ denotes the background (classical) metric 
corresponding to the line element of eqn.\ref{metric}. We 
focus our attention on the relevant (four-dimensional) 
tensor fluctuations $\Delta_{\mu \kappa}$ 
which, for the sake of convenience, are parametrized as
\begin{equation}
    \Delta_{\mu \kappa} = b^2(z) \, a^2(y) \, \td_{\mu \kappa}(x_\mu, y, z)
    \label{delta_tilde}
\end{equation}
The corresponding equation of motion is,
\begin{equation}
R_{\mu\kappa} = \frac{- \Lambda}{2} \, g_{\mu \kappa}
\end{equation}
The gauge conditions
\[
\Delta^\mu_\mu = 0 \ , \quad \qquad \partial^\mu \Delta_{\mu \kappa} = 0 \ ,
\]
in turn, imply
\begin{equation}
\td^\mu_\mu = 0 \ , \quad \qquad \partial^\mu \td_{\mu \kappa} = 0 \ .
\end{equation}
The KK mode expansion, in terms of the four-dimensional fields
$h^{(n,p)}_{\mu\nu}(x)$ can now be written in terms of the two 
winding numbers as
\begin{equation}
\td_{\mu\nu}(x^\mu, , z) = \frac{1}{\sqrt{R_y \, r_z}}
                            \sum_{n,p} h^{(n,p)}_{\mu\nu}(x) \,
                                    \psi_{np}(y) \, \chi_p(z) \ .
\end{equation}
This, then, yields the equations of motion, viz.
\begin{equation}
\barr{rcl}
0 & = & \dis (\Box + m^2_{np} ) \, h^{(n,p)}_{\mu\nu}(x)
\\[2ex]
0 & = & \dis R_y^{-2} \, \frac{d}{dy} \, 
                \left(a^4 \, \frac{d \psi_{np}}{dy} \right)
            - m_p^2 \, a^4 \, \psi_{np}
            + m_{np}^2 \, a^2 \, \psi_{np}
\\[2ex]
0 & = & \dis r_z^{-2} \, \frac{d}{dz} \, 
                \left(b^5 \, \frac{d \chi_p}{dz} \right)
            + m_p^2 \, b^3 \, \chi_p
\earr
\end{equation}
To obtain the spectrum, we need to solve the equations for the 
modes $\chi_p(z)$ and $\psi_{np}(y)$, which we now proceed to do.

\subsection{The $z$ equation}
For the zeroth mode, we have 
\[
\partial_z \, \left(b^5 \, \partial_z \, \chi_0 \right) = 0
\]
which has the particularly simple solution
\begin{equation}
\barr{rcl}
\chi_0 & = & \dis c_0^{(0)} + \frac{c_1^{(0)}}{8 \, k}  \, 
                \left[ 6 \, \tan^{-1} \left(\tanh \frac{k \, z}{2} \right)
                        + \left(3 + {\rm sech}^2 (k \, z) \right) \, 
                             {\rm sech} (k \, z) \, \tanh (k \, z) \right] \ .
\earr
     \label{chi_0}
\end{equation}
The constants $c_{0,1}^{(0)}$ are determined from the boundary conditions 
and/or normalization of the wavefunction $\chi_0(z)$. 
The solution for the higher modes $\chi_p$ are obtained in terms of  
associated Legendre polynomials of the 
first and second kinds, viz.
\begin{equation}
\barr{rcl}
\chi_p(z) & = & \dis \widetilde \Xi_p \, {\rm sech}^{5/2}(k \, z) \, 
      \left[ \cos\theta_p \, P_{\nu_p}^{5/2}\left(\tanh(k \, z) \right) 
           + \sin\theta_p \, Q_{\nu_p}^{5/2}\left(\tanh(k \, z) \right) \right]
\\[3ex]
\nu_p & \equiv & \dis \sqrt{4 + \frac{m_p^2 \,r_z^2 \, \cosh^2 (k \pi)}{k^2}}  
            - \, \frac{1}{2}
      = \sqrt{4 + \frac{m_p^2 \, R_y^2}{c^2} } \, - \, \frac{1}{2}
\\[2ex]
      & \equiv & \dis \sqrt{4 + x_p^2 \, \cosh^2 (k \pi) } \, - \, \frac{1}{2} 
\earr
     \label{chi_soln}
\end{equation}
where $\theta_p$ determines the relative 
weight of the two independent solutions
and $\widetilde \Xi_p$ is the normalization constant obtained from 
\begin{equation}
\barr{rcl}
\delta_{p \, p'} & = & \dis \int_{-\pi}^\pi dz \, b^3(z) \, \chi_p(z) \, 
                   \chi_{p'}(z) \ .
\earr
\end{equation}
That the above solution reduces to the aforementioned $\chi_0(p)$ for $m_p = 0$
(i.e., $\nu_p = 3/2$) is easy to see.

It should be noted that $\nu_p$ is not necessarily integral (or, even
half-integral). The presence of the associated Legendre functions
renders the analysis much more complicated than is the case for the 5D
analogue. This, in turn, introduces interesting new features. 

It has been argued in the literature~\cite{Mssg} that the $z$-equation
can be simplified to a great extent by approximating the warp factor
$1/\cosh(k \, z)$ by an exponential, which ought to be valid for large
$k \, z$. Indeed, thus simplified equation of motion has solutions in
terms of Bessel and Neumann functions, and the corresponding analysis
has exact parallels with the 5D case. The approximation however would
not work for the small $k$ regime. Moreover even for large $k$, such
approximation is invalid for $z \sim 0$, precisely the region
where we are supposed to be located. And since the values of the
graviton wavefunctions would determine the strength of their couplings
to the SM fields, we should expect that such an approximation would
lead to some inaccuracies.
Moreover, such an approximation changes the differentiability 
of the warp factors, thereby changing the boundary conditions on the 
graviton wavefunctions. 
As we shall see later, the consequences of such an 
approximation are really profound and, hence, we desist from adopting it.

\subsection{The $y$ equation}
The equation for the $y$-mode function can be simplified 
by making the transformations
\begin{equation}
\barr{rcl}
\psi_{np}(y) & = & \dis e^{2 \, c \, |y|} \, \bar \psi_{np}(\theta)
\\[3ex]
\theta & = & \dis  \frac{m_{np} \, R_y}{c} \, e^{c \, |y|} \ ,
\earr 
\end{equation}
leading to
\[
 \theta \, \frac{d \bar \psi_{np}}{d \theta}
	         + \theta^2 \, \frac{d^2 \bar \psi_{np}}{d \theta^2}
- \left( 4 + \frac{m_p^2 \, R_y^2}{c^2} \right) \, \bar \psi_{np} 
            +  \theta^2 \, \bar \psi_{np}
= 0
\]
This, again, leads to a solution in terms of Bessel functions of the
first and second kinds, viz.
\begin{equation}
\psi_{np}(y)  =  \Xi_{np} \, e^{2 \, c \, |y|} \, 
            \left[J_{\nu_p + \frac{1}{2}}(\theta) 
                + \zeta_{np} \, Y_{\nu_p + \frac{1}{2}}(\theta) \right] \ ,
\end{equation}
where $\nu_p$ has been defined earlier. Once again, the constants $\Xi_{np}$ 
and $\zeta_{np}$ are to be determined by using the orthonormality conditions, 
viz.
\begin{equation}
\barr{rcl}
\delta_{n \, n'} & = & \dis \int_{-\pi}^\pi dy \, a^2(y) \, \psi_{np}(y) \, 
                   \psi_{n'p}(y) 
\earr
\end{equation}
The parallel with the 5D case is very apparent and, thus, all the
analyses for the original RS case can be trivially transported to this
sector. However, it should be appreciated that $\psi_{np}$ are
crucially dependent on the eigenspectrum of the $z$-equation
operator. Indeed, the very order of the Bessel functions ($\nu_p +
1/2$) is determined entirely by it. While this may, at first, seem to
imply that the spectrum is determined by a single parameter $\nu_p$,
note that it is not so, for the others enter through $\theta$. A
further issue needs to be clarified here. It has been 
argued in the literature (\cite{Mssg} as well as in the context of a
different system with close parallels to the current discussion) that,
for $p \neq 0$ modes such as $\psi_{0 p}$ would not exist. We shall
explicitly show below that this is not the case.

\subsection{Mass spectrum for the KK graviton}

Our aim, now, is to compute the allowed values of $m_{np}$ (i.e., the
KK graviton masses). We first obtain these in terms of $m_{p}$, the
eigenvalues of the $z$-direction differential operator, and, then,
determine $m_p$. Either exercise is crucially dependent on the
differentiability structure of the wavefunctions.

The self-adjoint nature of the $y$-direction operator implies that the 
derivatives $\psi'_{np}(y)$ must be continuous at either boundary. Note that 
the presence of the brane tension has, in essence, been factored out 
by the inclusion of the warp factors in the definition of $\td_{\mu \kappa}$
(see eqn.\ref{delta_tilde}). This, then, leads to 
\begin{equation}
\left. \zeta_{np}
= \, - \, \frac{ x_{np} \, e^{c \, (|y| \, - \, \pi)} 
             J_{\nu_p -\frac{1}{2}}(x_{np} e^{c \, (|y| \, - \, \pi)})
        +(\frac{3}{2}- \nu_p) J_{\nu_p +\frac{1}{2}}(x_{np} e^{c \, (|y| \, - \, \pi)})}
       { x_{np} e^{c \, (|y| \, - \, \pi)} 
        Y_{\nu_p -\frac{1}{2}}(x_{np} e^{c \, (|y| \, - \, \pi)})
+(\frac{3}{2} - \nu_p) Y_{\nu_p +\frac{1}{2}}(x_{np} e^{c \, (|y| \, - \, \pi)})}
\right|_{y = 0, \pi}
   \label{ycond}
\end{equation}
where
\begin{equation}
x_{np} \equiv m_{np} \, \frac{R_y}{c} \, e^{c \, \pi} \ ,
\end{equation}
and the two conditions summarised in eqn.(\ref{ycond} reflect 
the boundary conditions at $y = 0, \pi$ respectively.
Once $\nu_p$ is known, 
these two, together, 
determine $\psi_n(y)$ as well as serve to quantize 
$x_{np}$ (and, hence, $m_{np}$).

We now turn our attention to $\chi_p(z)$. 
As these have to be even functions of $z$, we have 
$\chi_p'(z = 0) = 0$. This is identically satisfied by $\chi_0(z)$ as
$\nu_p(m_p = 0) = 3/2$ and the corresponding functions satisfy
$P_{3/2}^{5/2}(x) \propto (1 - x^2)^{-5/4}$
and $Q_{3/2}^{5/2}(x) = 0$. For $p \neq 0$, we may use 
the identities
\[
\barr{rcl}
\dis \left( \frac{d P_N^M (x)}{d x} \right)_{x = 0}
  & = & \dis 
\frac{2^{M + 1}}{\sqrt{\pi}} \,
    \sin\left( \frac{\pi \, (N + M)}{2} \right) \,
    \frac{\Gamma( 1 + (N + M) / 2)} {\Gamma( (N - M + 1) / 2)}
\\[3ex]
\dis \left( \frac{d Q_N^M (x)}{d x} \right)_{x = 0}
  & = & \dis 
2^{M} \, \sqrt{\pi} \,
    \cos\left( \frac{\pi \, (N + M)}{2} \right) \,
    \frac{\Gamma( 1 + (N + M) / 2)} {\Gamma( (N - M + 1) / 2)}
\earr
\]
leading to
\begin{equation}
\cot \theta_p = \frac{-\pi}{2} \, \cot \frac{\pi \, (\nu_p + 5/2)}{2} \ .
\end{equation}

To determine the mass spectra of the KK gravitons, we need to analyze
the continuity condition at $z = \pi$ which, for convenience, we
separately consider in two distinct cases namely large and small $k$.

\vskip 10pt
\noindent
\underline{\bf Large $k$ (small $c$)}

Denoting $\tau = \tanh(k \, z)$, we have
\[
\chi_p(z)  =  \dis \widetilde \Xi_p \, (1 - \tau^2)^{5/4} \, 
      \left[ \cot\theta_p \, P_{\nu_p}^{5/2}(\tau)
           + \, Q_{\nu_p}^{5/2}(\tau) \right] \ .
\]
As the orbifolding condition necessitates\footnote{Since
$\chi_p(z)$ is even, its derivative $f(\tau)$ is odd. On the other
hand, the orbifolding and the continuity of the derivative imply
$\chi'_p(z= \pi_-) = \chi'_p(z= \pi_+) = \chi'_p(z= -\pi_-)$.} that 
$\chi'_p(z = \pi) = 0$, we need to examine the derivative close to 
$\tau = 1$. For the zero mode ($\nu_p = 3/2$ or $m_p = 0$) this implies
$c_1^{(0)} = 0$ in eqn.(\ref{chi_0}), or in other words, $\chi_0(z)$ 
is flat (as would be expected). 
For the others, we have
\[
f(\tau) \equiv \frac{d \chi_p}{d \tau} 
        =  
\frac{\widetilde \Xi_p}{2} \, (2 \nu_p -3) \sqrt[4]{1-\tau^2} 
\left[\cot\theta_p \, \tau P_{\nu_p \
}^{5/2}(\tau)-\cot\theta_p\, P_{\nu_p +1}^{5/2}(\tau)+\tau Q_{\nu_p \
}^{5/2}(\tau)-Q_{\nu_p +1}^{5/2}(\tau)\right] \ .
\]
In the infinitesimal neighbourhood of $\tau = 1$,
\[
   f(\tau = 1 - \delta) = \
   \frac{\cot\theta_p \, (2 \nu_p -3) (2 \nu_p +5)}{2 \sqrt{2 \pi }} \,
   \left[ - 1 + \delta \, 
\frac{(2 \nu_p -1) (2 \nu_p +3)}{4} \right] + {\cal O}(\delta^2) \ .
\]
For the higher modes ($\nu_p > 3/2$), the disappearance of
$\chi'_p(z = \pi)$, thus, needs $\cot\theta_p = 0$ or
\begin{equation}
   \nu_p =  2 \, n + \frac{1}{2} \qquad n \in Z^+
    \label{nup_in_large_k}
\end{equation}
This result can be appreciated by noting that $P_{\nu_p > 3/2}^{5/2}(\tau) 
\to \infty$ as $\tau \to \pm 1$. Since, for large $k$, the 
wavefunctions must extend close to $\tau \approx \pm 1$, 
normalizability of the same requires $\cot\theta_p \to 0$.

Using eqn.(\ref{nup_in_large_k}) in the second of eqns.(\ref{chi_soln}) 
would determine the allowed values of $m_p$. Substituting the latter 
in  eqn.(\ref{ycond}) would, then, yield the allowed values of $m_{n,p}$,
or, in other words, the spectrum. However, since a large $k$ 
implies a $c$ that is almost infinitesimally small, there is virtually 
no warping in the $y$--direction and the latter is essentially flat. 
This would immediately imply that $m_{np}^2 \approx m_p^2 + n^2 \, R_y^{-2}$. 
With $R_y$ being very small, $h^{(n > 0, p)}$ are too heavy to be 
of any relevance, and we effectively have but a single tower 
$h^{(0,p)}$ with masses $m_{0p} \approx m_{p}$.

\vskip 10pt
\noindent
\underline{\bf Small $k$ (Large $c$)}

The boundary is now at $\tau = \tau_\pi \, = \, \tanh( k \, \pi)$, and
somewhat away from $\tau = 1$.  Being away from the 
singular points of the associated Legendre functions means one 
can numerically calculate the functions, and the vanishing 
of $f(\tau_\pi)$ dictates that 
\begin{equation}
\cot\theta_p \, \tau_\pi P_{\nu_p \
}^{5/2}(\tau_\pi)-\cot\theta_p\, P_{\nu_p +1}^{5/2}(\tau_\pi)+\tau_\pi Q_{\nu_p \
}^{5/2}(\tau_\pi)-Q_{\nu_p +1}^{5/2}(\tau_\pi) \, = \, 0 \ .
\end{equation}
This equation has to be solved numerically to obtain the quantized values
of $\nu_p$. To now obtain $x_{np}$, concentrate on eqn.(\ref{ycond}). 
Since $e^{c \, \pi} \gg 1$, this relation is satisfied only if
\begin{equation} 
\dis 2 x_{np} J_{\nu_p -\frac{1}{2}}(x_{np} ) +(3-2 \nu_p )
J_{\nu_p +\frac{1}{2}}(x_{np} ) \, = \, 0 \ .
\end{equation}
Finally, for large $c$, the graviton spectrum will be given by the solutions
of the above equation. It is worth remembering that, in this case, there 
is a non-negligible warping in the $z$-direction, and thus, the 
$h^{(n, p > 0)}$ are not necessarily superheavy. The two branches (large $k$ 
and large $c$) are, thus, not quite symmetrical. 

\subsection{Couplings with brane fields}

The interaction term of a graviton with any brane field is 
given by 
\begin{equation} 
L_{\rm
int}=\frac{1}{M_6^{2}} T^{\mu \nu}h_{\mu \nu}(x_{\mu}, y=\pi,z=0) \ ,
\end{equation} 
where $T^{\mu \nu}$ is the energy-momentum tensor of the field. 
The coupling of brane-localized matter with the $(n,p)^{th}$ graviton mode
is, thus, determined by the value of the latter's wavefunction on the 
brane location. In other words, 
\begin{equation}
\label{coup}
C_{np} = \frac{1}{M_6^{2}\sqrt{R_y r_z}} \, \Psi_{np}(\pi) \, \chi_{p}(0) \ .
\end{equation}
Once again, we examine the two cases separately.

\vskip 10pt
\noindent
\underline{\bf Large $k$ (small $c$)}

In this case, as argued earlier, the lowest mass modes correspond to
the $\psi_{0p}$ states.  From the solutions of $\psi_{np}(y)$ and
$\chi_{p}(z)$, we have
\[
\psi_{0p}(\pi)  =  \Xi_{0p} \ , 
\qquad 
\chi_{p = 0}(0) =  \widetilde\Xi_0 \ , 
\qquad
\chi_{p \, \neq \, 0}(0)  = \widetilde\Xi_p \, 
      \left[ Q_{\nu_p}^{5/2}\left(0 \right) \right] \ ,
\]
where $\Xi_{0p}$ and $ \widetilde\Xi_{p}$ are to be determined from the
orthonormality conditions of the mode functions.  From eq(\ref{coup})
we then have
\begin{equation}
\barr{rcl}
C_{00} 
& = & \dis
 \frac{1}{M_6^{2}\sqrt{2 \pi \, R_y r_z}} \,  
       B_0^{-1/2} \, \cosh^{3/2}(k\pi)
\\[5ex]
C_{0p} & = & \dis
\frac{1}{M_6^{2}\sqrt{2 \pi \, R_y r_z}} \,  
B_p^{-1/2} \, \cosh^{3/2}(k \pi) \,
\left[ Q_{\nu_p}^{5/2}\left(0 \right) \right] \ ,
\earr
\end{equation}
where
\begin{equation}
\barr{rcl}
\\[3ex]
B_{p \, = \, 0} & \equiv &  \dis \int_{-\pi}^{\pi} \cosh^3(k \, z)\,  dz
\\[3ex]
B_{p \, \neq \, 0} & \equiv & \dis \int_{-\pi}^{\pi} {\rm sech}^2(k \, z) 
\left[Q_{\nu_p}^{5/2}(\tanh(k \, z)) \right]^2 dz \ .
\earr
\end{equation}

In the above, terms subleading in $c$ have been dropped as $c \ll 1$.

\vskip 20pt
\noindent 
\underline{\bf Small $k$ (large $c$)}

In this case, the wavefunctions on our brane are given by
\[
\psi_{np}(\pi)  =  \Xi_{np} \, e^{2 \, c \, \pi} \,
            J_{\nu_p + \frac{1}{2}}(\theta_\pi) \ ,
\qquad 
\chi_{p = 0}(0)  =  \widetilde \Xi_0 \ ,
\qquad
\chi_{p > 0}(0) = \widetilde \Xi_p \, 
      \left[ \cot\theta_p \, P_{\nu_p}^{5/2}(0)
           + \, Q_{\nu_p}^{5/2}(0) \right] \ .
\]
As before, $\Xi_{np}$ and $ \widetilde \Xi_{p}$ are to be determined
from the normalizations.  Once again, to determine the couplings we
refer to eqn.(\ref{coup}) which yields
\begin{equation}
\barr{rcl}
C_{n0} 
& = & \dis
  \frac{1}{M_6^{2}r_z} \, \cosh(k \, \pi)\, e^{c \, \pi} \sqrt{\frac{k}{2 \, A_{n0} \, B_0}} \,
            \left[ J_{2}(\theta_\pi) \right]
\\[5ex]
C_{n,{p\neq0}} &= & \dis
 \frac{1}{M_6^{2}r_z} \,  \cosh(k \, \pi)\, e^{c \, \pi} \sqrt{\frac{k}{2 \, A_{np} \, B_p}} \,
            \left[ J_{\nu_p + \frac{1}{2}}(\theta_\pi) \right] \left[ \cot\theta_p \, P_{\nu_p}^{5/2}(0)
           + \, Q_{\nu_p}^{5/2}(0) \right] \ ,
\earr
\end{equation}
where,
\begin{equation}
\barr{rcl}
A_{np} & = & \dis \int_{0}^{1} r \, 
\left[J_{\nu_p +\frac{1}{2}}(x_{np} \, r)\right]^2 dr
\\[3ex]
B_{p \, = \, 0} & = & \dis \int_{-\pi}^{\pi} \cosh^3(k \, z)\, dz
\\[3ex]
B_{p \, \neq \, 0} & = &
 \dis \int_{-\pi}^{\pi} {\rm sech}(k \, z)^2 \left[ \cot\theta_p \, P_{\nu_p}^{5/2}(\tanh(k \, z))
           + \, Q_{\nu_p}^{5/2}(\tanh(k \, z)) \right]^2 dz \ .
\earr
\end{equation}

Several points need to be noted at this point.
\begin{itemize}
\item 
Unlike in the previous case, the KK-modes in the $y$-direction are 
now relatively light and visible. This is but a
consequence of the fact that the $y$-direction warping is dominant.
\item 
Although the $z$-warping is subdominant, it is not entirely
  negligible. (This is quite contrary to the other case, where the
  $y$-warping was virtually nonexistent.) Thus, there is hope that
  some of the $z$-direction modes might be visible.
\item 
In addition, the wave function in $y$-direction is dependent on $p$ 
(the momentum in the $z$-direction). 
\item
For $p=0$, the levels $h^{(n,0)}$ have almost the same coupling with the 
SM fields for $n > 0$. While this may seem counterintuitive given that 
the normalizations $A_{n0}$ depend on $n$, the same is essentially cancelled 
by the $n$-dependence in $J_2(\theta_\pi)$. Indeed, this result is 
exactly analogous to that for the (five-dimensional) RS model, and was 
to be expected given that the $h^{(n,0)}$ wavefunctions 
are flat in the $z$-direction. 
On the other hand, for a given $p > 0$, increasing 
$n$ results in the suppression of the corresponding couplings. Understandably,
the extent of this suppression increases with $k$ (which is a measure 
of the subdominant warping).
\item 
For the very same reason, increasing $p$, while keeping $n$ constant leads 
to an enhancement of the couplings. 
\end{itemize}

\section{Numerical values for masses and couplings}

In exploring the parameter space of the model, it is useful to consider 
two dimensionless quantities
\begin{equation}
\epsilon \equiv \frac{k}{r_z \, M_6}  \ , \quad
\alpha \equiv \frac{R_y}{r_z} \ .
\end{equation}
Quite analogous to the 5D case, here too the
applicability of the classical solutions can be related to the 
issue of the bulk curvature being small sufficiently small compared to
$M_6$. To this end, we shall demand that $\epsilon < 0.1$. On the 
other hand, we would not like to introduce a new hierarchy (between moduli) 
in our efforts to ameliorate the SM hierachy problem. Thus, the ratio 
$\alpha$ should neither be too large nor too small. 

We can, then, explore the parameter space of the theory in terms of 
$\epsilon, \alpha$ and any one other, say $M_6$ (or, equivalently $k$), 
relating all the rest 
through eqns.(\ref{RS6_eqns}\&\ref{Planck_mass}). A very important 
distinction from the original RS scenario is that $M_6$ need not be 
nearly the same as the four-dimensional Planck mass $M_P$. This freedom 
accrues from the larger parameter space of the model. In fact, 
$M_6$ can be significantly smaller than $M_P$ without any fine tuning. 
Indeed, the large $c$ branch needs $\alpha \gapp 50$ and with 
\[
   M_P^2 \sim \frac{M_6^4 \, r_z \, R_y}{2 \, c \, k}  \,
         = \frac{M_6^4 \, r_z^2 \, \alpha}{2 \, c \, k} 
\]
even the largest allowed value of $k$ ($\lapp 0.3$) would lead to 
$M_6 \lapp M_P / 2$. Smaller (larger) values of $k$ ($\alpha$) 
would lead to even smaller $M_6$. 

Furthermore, in this scenario, the cutoff for a four-dimensional
quantum field theory is set not by $M_6$, but by $\mbox{min}(R_y^{-1},
r_z^{-1})$.   At such a scale,
  the higher-dimensional nature of the theory becomes apparent, and
  the four-dimensional effective theory (including the graviton modes)
  is no longer an apt language to describe physics.\footnote{A
      parallel is provided by an ADD-like~\cite{arkani} model with
      unequal radii of compactification. In fact, in the bulk, the large-$k$ 
      branch is conformal to $RS_5\otimes ADD$, with the correspondence 
      broken only by the brane tensions.} 
Indeed, while the mechanism of compactification cannot be addressed 
in our theory (or within the RS mechanism), the physics responsible for 
it must be taken into account in any description that reaches beyond 
this scale. In other words, the
quantity $w^{-1}$ as defined in eqn.(\ref{hierarchy}) refers to the
ratio of the Higgs mass and this cutoff scale and is no longer
constrained to be $\gapp 10^{16}$.  Indeed, it can be significantly
smaller.  Once again, this freedom (absent in the 5D analogue) is but
a consequence of the larger parameter space of the present theory.

At this stage, we wish to clarify an issue regarding effective
theories that, often, leads to miscommunication. The cutoff scale of
an effective theory is often described as the scale at which the loop
contributions (often very large) are to be cutoff, for the new physics
beyond this scale would naturally regulate them ({\em i.e.} cancel
unwanted divergences).  However, for this cancellation to be
demonstrated, the said ultraviolet completion has to be known
exactly. This is certainly not the case here (quite unlike, say the
MSSM or gauge-Higgs unification scenarios, wherein the amelioration of
the large corrections can be shown explicitly). On the contrary, {\em
  sans} a reliable theory of quantum gravity, no such calculation is
possible. It has been argued that, within the five-dimensional
context, the addition of the Planck-brane and/or the TeV-brane allows
a holographic interpretation~\cite{malda}, with the former acting as a
regulator leading to a UV cutoff ($\lapp r_c^{-1}$) on the
corresponding
CFT~\cite{ArkaniHamed:2000ds,Rattazzi:2000hs,PerezVictoria:2001pa}.
Similar analyses have also been made for theories with gauge fields
extended in to the warped
bulk~\cite{Pomarol:1999ad,Rizzo,Agashe:2002jx}. Although no such
duality has been constructed for the six-dimensional case, it is quite
conceivable that one such would exist (for the large $k$ case, the
bulk is indeed AdS$_6$-like). Consequently, even on this count, the
branes are expected to provide a regulator with a cutoff $\lapp
\mbox{min}(R_y^{-1}, r_z^{-1})$.In particular, let
  us concentrate on the situation $R_y > r_z$, which is mostly the
  case (with exceptions to this generically being bad
  phenomenologically). Remembering that the space is orbifolded on
  $S^{1}/Z_2\otimes S^{1}/Z_2$, let us concentrate on the 4-brane at
  $z = 0$ (with us being localized at the $z = 0, y= \pi$
  intersection). This 4-brane, thus, reflects a $AdS_5$ geometry in
  the bulk. Indeed, viewed in isolation, it is but a perturbation of
  the $RS\, 1$ scenario with a corresponding CFT cutoff of
  $R_y^{-1}$. Thus, this part of the parameter space is manifestly
  consistent with our assertion about the cutoff.

We now examine the allowed parameter space in the light of the
discussion above, considering, in turn, the large $c$ and large $k$
cases.

\subsection{Small $k$ (large $c$)}
In Table \ref{tab:small_k},
we present part of the spectra for four
representative points in the parameter 
space, each corresponding to a 
particular value of the ratio of the bulk curvature and the 
quantum gravity scale, namely $\epsilon = 0.0775$.
Once  $\epsilon$ is fixed, for this branch of the solution, $c$  has only 
a very subdominant dependence on $k$ (see eqn.(\ref{hierarchy})). 
The relation $c = \alpha k / \cosh(k \pi)$ would, then, imply that 
a larger $k$ needs a smaller $\alpha$, as is demonstrated 
by Table \ref{tab:small_k}. On the other hand, since the 
modes $h^{(n,0)}$ are flat in the $z$-direction, the masses $m_{n0}$ 
are essentially free of $k$,xs with the small difference in 
Table \ref{tab:small_k} accruing from the difference in the values 
of the other parameters. 

\input{Tab_small_k.tex}

The masses $m_{n1}$, on the other hand, do exhibit considerable dependence on 
$k$. Moreover, these modes are considerably heavier than several of 
the $h^{(n,0)}$. As can be expected, these masses grow very fast as
$k$ becomes smaller, a consequence of the decreasing severity of the 
$z$-warping.

What is of particular significance in each case is that the masses 
are much larger than what has been probed at the LHC. Indeed, masses 
such as these were practically out of reach of the runs at 
$\sqrt{s} = 7, 8 \tev$, and would be accessible only in the next run. 
However, with the couplings to the SM fields being much smaller than those 
for the original RS gravitons, the production rates would continue 
to be highly suppressed even at the future runs at $\sqrt{s} = 13, 14 \tev$.
Indeed, as Table \ref{tab:small_k} suggests, 
for the large $c$ branch of the solution, discovering 
even the first graviton mode at the LHC will remain a dream unless 
$k$ is very small indeed, when the system becomes RS-like with the 
graviton couplings increasing appropriately. On the other hand, such 
values of $k$ typically necessitate a somewhat large value of $\alpha$.

\begin{figure}[!h]
{
\vspace*{-10pt}
\hspace*{-40pt}
\epsfxsize=6cm\epsfbox{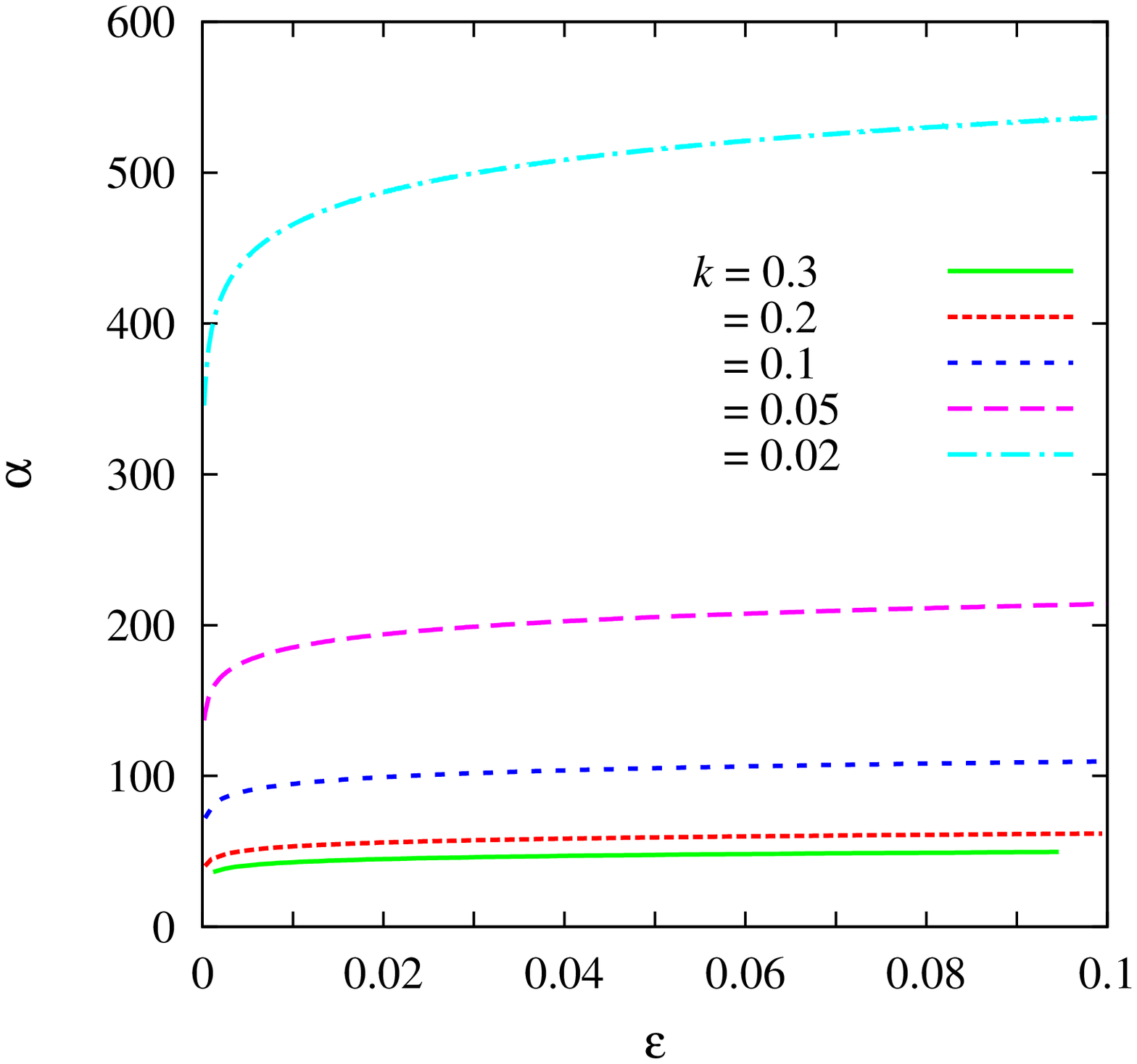}
\epsfxsize=6cm\epsfbox{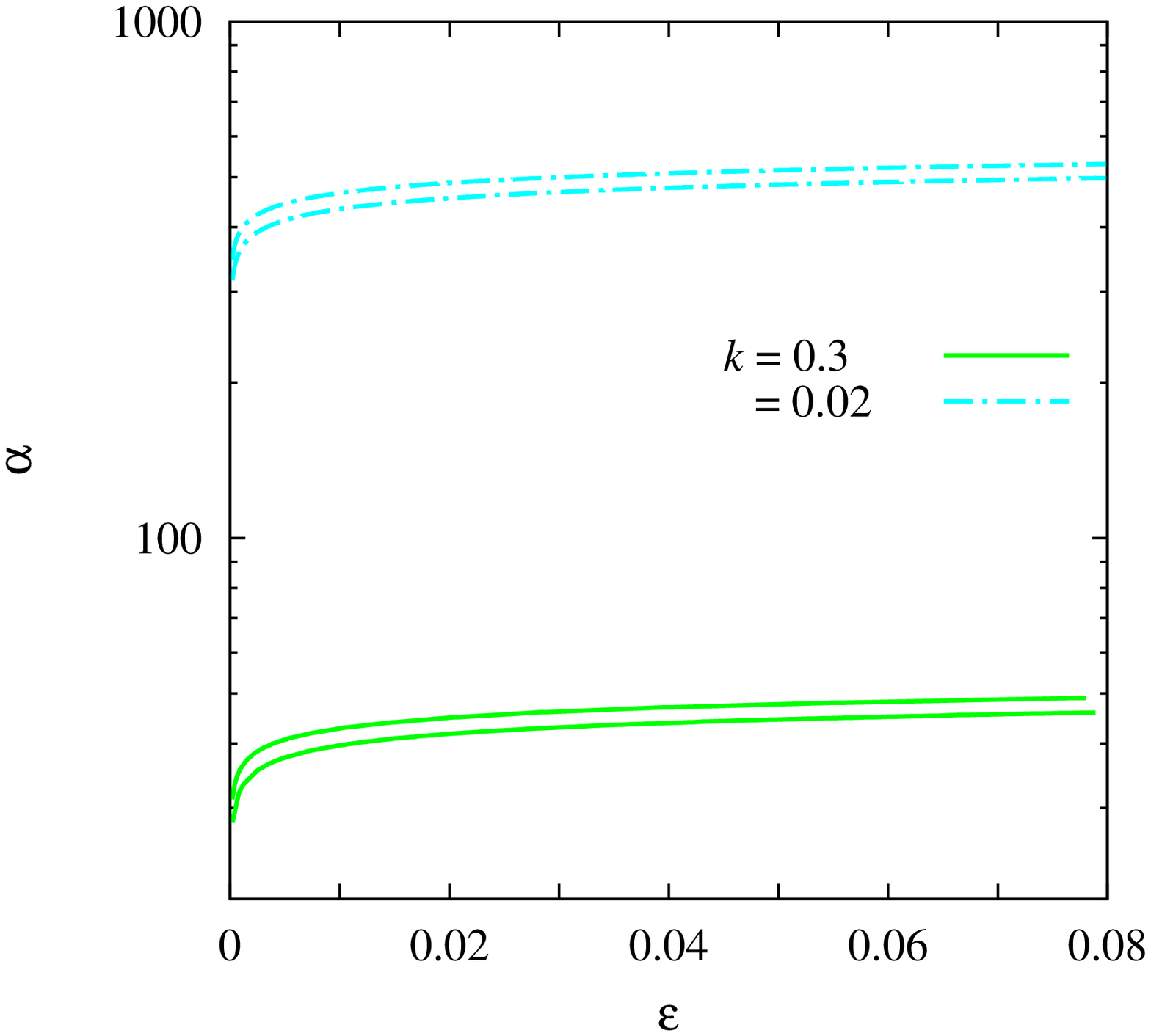}
\vspace*{-60pt}
}
\caption{(Left panel) Contour plots in the $(\epsilon, \alpha)$ plane
  for fixed values of $k$. The curves are constrained to satisfy $w \,
  R_y^{-1} = m_h $. (Right panel) The dependence of the contours on
  the value of the ratio $w / R_y$. In each case, the upper and lower
  curves correspond to $m_h$ and $1 \tev$ respectively.}
\label{fig:small_k_alpha_eps}
\end{figure}

Such conclusions are brought into focus by
Fig.\ref{fig:small_k_alpha_eps} where we have depicted the relation
between the parameters $(\alpha, \epsilon)$ that, for a given choice
of $k$, leads to the correct hierarchy (with the ultraviolet cutoff
being given by $R_y^{-1}$). The modulus ratio $\alpha$ is a
monotonically increasing (decreasing) function of $\epsilon$ ($k$),
with the dependence on $k$ being much more pronounced. In the
  left panel of the figure (as also in the subsequent numerical
  analysis), we hold $m_h = w \, M_{\rm cutoff}$ with the cutoff scale
  being defined by the larger of the two compactification radii. While
  this choice of the hierarchy factor $w$ is certainly as good as any
  other, the numerical results are not greatly sensitive to the exact
  value. This is borne out by the right panel of the same figure,
  which demonstrates (for the two extreme choices of $k$ in the left
  panel) that the values remain qualititatively the same even if we
  change $w$ by a factor of 8.

\begin{figure}[!h]
{
\vspace*{-10pt}
\hspace*{-40pt}
\epsfxsize=6cm\epsfbox{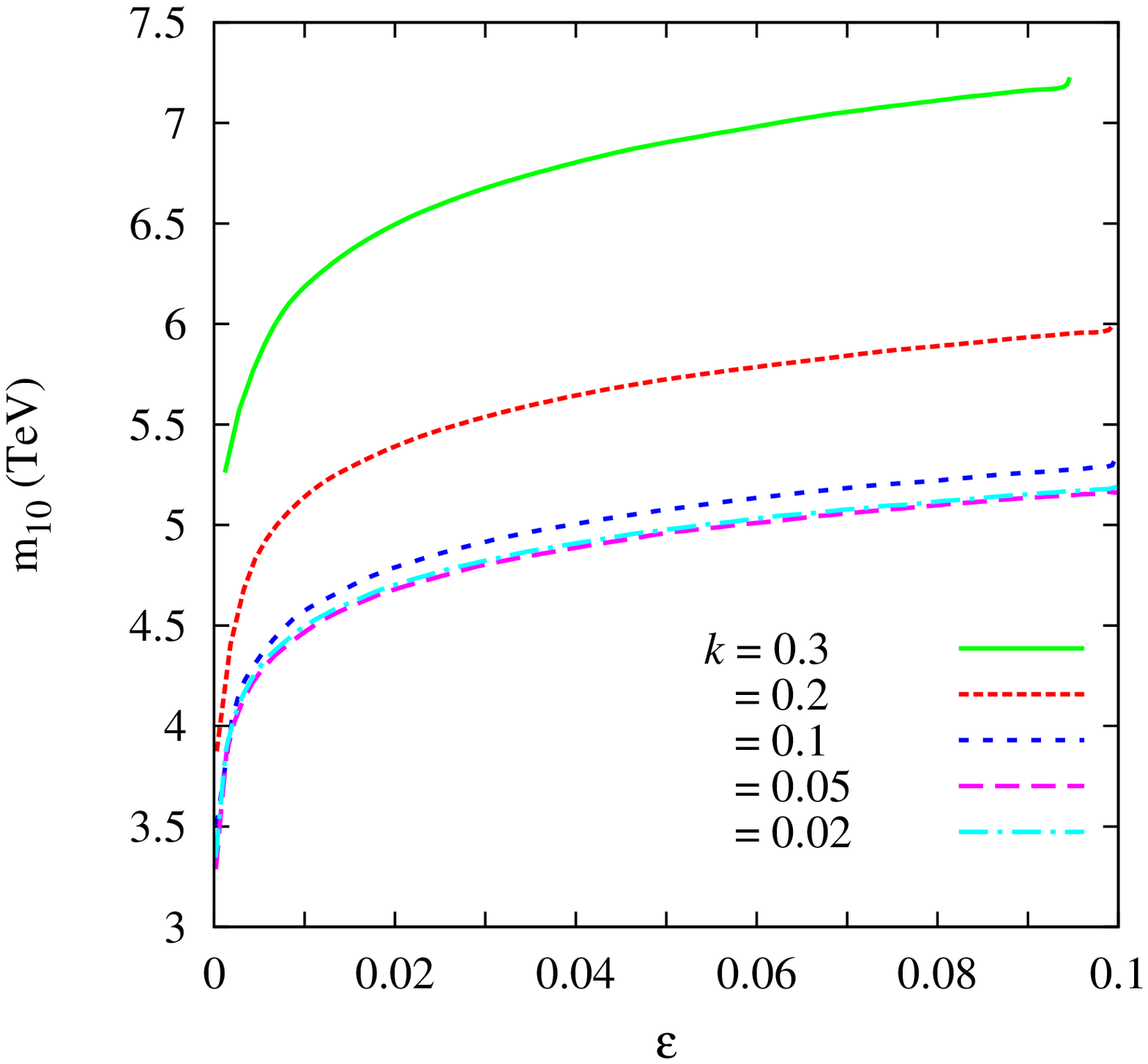}
\epsfxsize=6cm\epsfbox{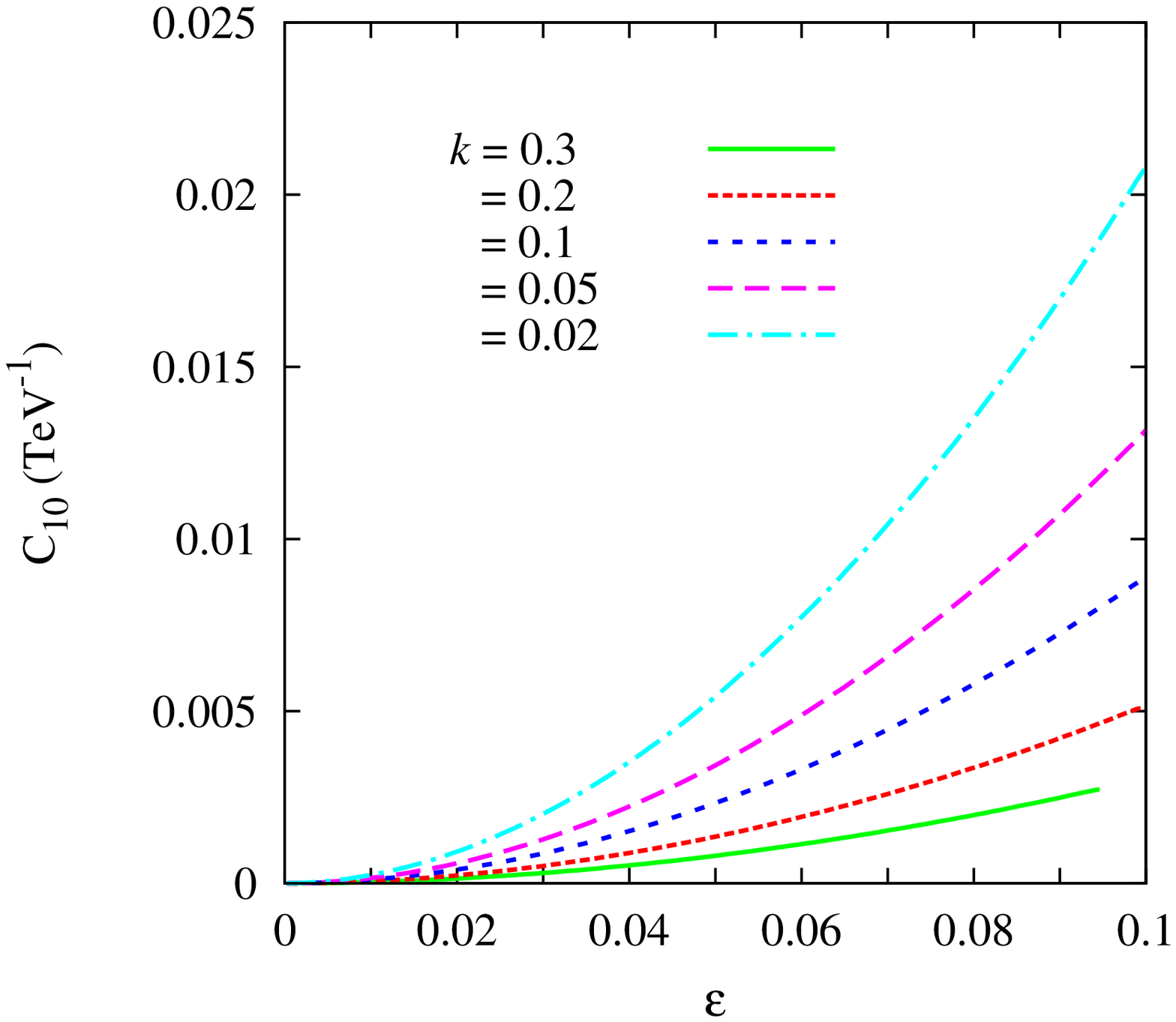}

\vspace*{-60pt}
}
\caption{The mass $ m_{10}$ (left panel) and matter coupling $C_{10}$ (right panel)
for the first graviton mode as a function of $\epsilon$ for a fixed $k$.  
The parameter $\alpha$ has been constrained to satisfy
$w \, R_y^{-1} = m_h $.}
\label{fig:small_k_mass&coup}
\end{figure}

In Fig.\ref{fig:small_k_mass&coup}, we depict the 
corresponding mass and SM-coupling strength of the lowest non-trivial 
graviton, viz. $h^{(1,0)}$. As has been argued above, decreasing $k$ not 
only makes this graviton lighter, but also strengthens its couplings, thereby 
making it more amenable to discovery at the LHC. This trend holds for the 
other modes too. 
The existence of the double tower is another interesting point 
to note, especially for not too small values of $k$. As 
Table \ref{tab:small_k} shows, one can have a clustering 
of the KK modes, each of which has an enhanced coupling to
the SM fields, and are likely to be seen in future experiments 
as a series of relatively closely lying resonances, with 
almost identical decay patterns. This proliferation of 
KK modes will be further enhanced if
the number of extra dimensions increases.

\subsection{Large $k$ (small $c$)}

The situation changes considerably now when compared to the preceding case.
With $c$ being very small, the low-lying spectrum is essentially independent 
of it. And, as alteady stated, with the $y$-direction suffering virtually 
no warping, all $h^{(n,p)}$ are superheavy ($ m_{np} > R_y^{-1}$) 
for $n > 0$, and, henceforth, we shall concentrate only on $h^{(0,p)}$. 

\input{Tab_large_k_1.tex}

As Table \ref{tab:large_k_1} shows, $\alpha$ can be much smaller now
(even smaller than one), and a large hierarchy between the moduli is
no longer necessary. Indeed, the smaller $\alpha$ is, the lighter the
graviton excitations are. The dependence of the masses on $k$ is
subdominant, though. These two features can be understood by recalling
that the masses, in this case, are essentially given by $m_p$, the
eigenvalues of the $z$-equation of motion. If we had a flat
$z$-direction, the eigenvalues would have been evenly separated,
namely $m_p = p / r_z$.  In the current scenario, this is tempered by
the warping. Since, for large $k$, the hierarchy is almost uniquely
determined ($w \approx {\rm sech} (k \, \pi)$), so is the cutoff scale
$R_y^{-1}$. Consequently, a smaller $\alpha$ implies a smaller
$r_z^{-1}$ and, hence, a lighter spectrum. If $M_6$ were to be held
constant, this would also translate to a smaller $\epsilon$, as hinted
at by Table \ref{tab:large_k_1}. The dependence of the masses on $k$,
thus, accrues, only through the warping and unless the latter changes 
by a great degree, the former remain relatively stable. 

The arguments above also tell us why the couplings $C_{0p}$ are
insensitive to $\alpha$. With the $h^{(0,p)}$ wavefunctions being
independent of $y$, any dependence of the couplings on the parameters
of the $y$-equation must disappear. Note, though, that the couplings
of the gravitons to the SM fields are much larger now than was the
case for the small $k$ branch of the theory. In fact, $C_{0p}$ for the 
two $k = 8.2$ points listed in  Table \ref{tab:large_k_1} are of the same 
order of magnitude as those for the RS model as investigated by 
the ATLAS collaboration~\cite{ATLAS1}.
Consequently, the gravitons for $k = 8.2, \alpha = 1.56$ should definitely 
be visible as resonances in the next run of the LHC, while those 
corresponding to $k = 8.2, \alpha = 9.87$ may leave behind some 
indications through virtual diagrams (at least in the high luminosity 
run).

Things take a more interesting turn for larger $k$ values, as the
couplings increase substantially (the entries on the right colum of
Table \ref{tab:large_k_1}). While the $k = 8.5, \alpha = 1.56$
gravitons would be seen as very prominent resonances, even the large
contact interactions generated by the $k = 8.5, \alpha = 9.87$ would
alter the continuum spectrum for the associated processes to a
significant degree. If we increase $k$ even further (see Table
\ref{tab:large_k_2}), the couplings rise very fast and quickly cross
over to the nonperturbative regime. This is but a consequence of the
fact that the wavefunctions $\chi_p(z)$ are highly concentrated near
$z = 0$ with the extent of peaking increasing with $k$. This hitherto
undiscovered strongly-coupled sector of the theory is potentially of
great theoretical interest. The strong coupling, though, does not
manifest itself for $k \lapp 9.0$ and a perturbative treatment does
make sense. In summary, the parameter region corresponding to $k \lapp
8.5$ is still far from being ruled out and admits very interesting
phenomenology.

\input{Tab_large_k_2.tex}

\begin{figure}[!h]
{
\vspace*{-10pt}
\hspace*{-40pt}
\epsfxsize=6cm\epsfbox{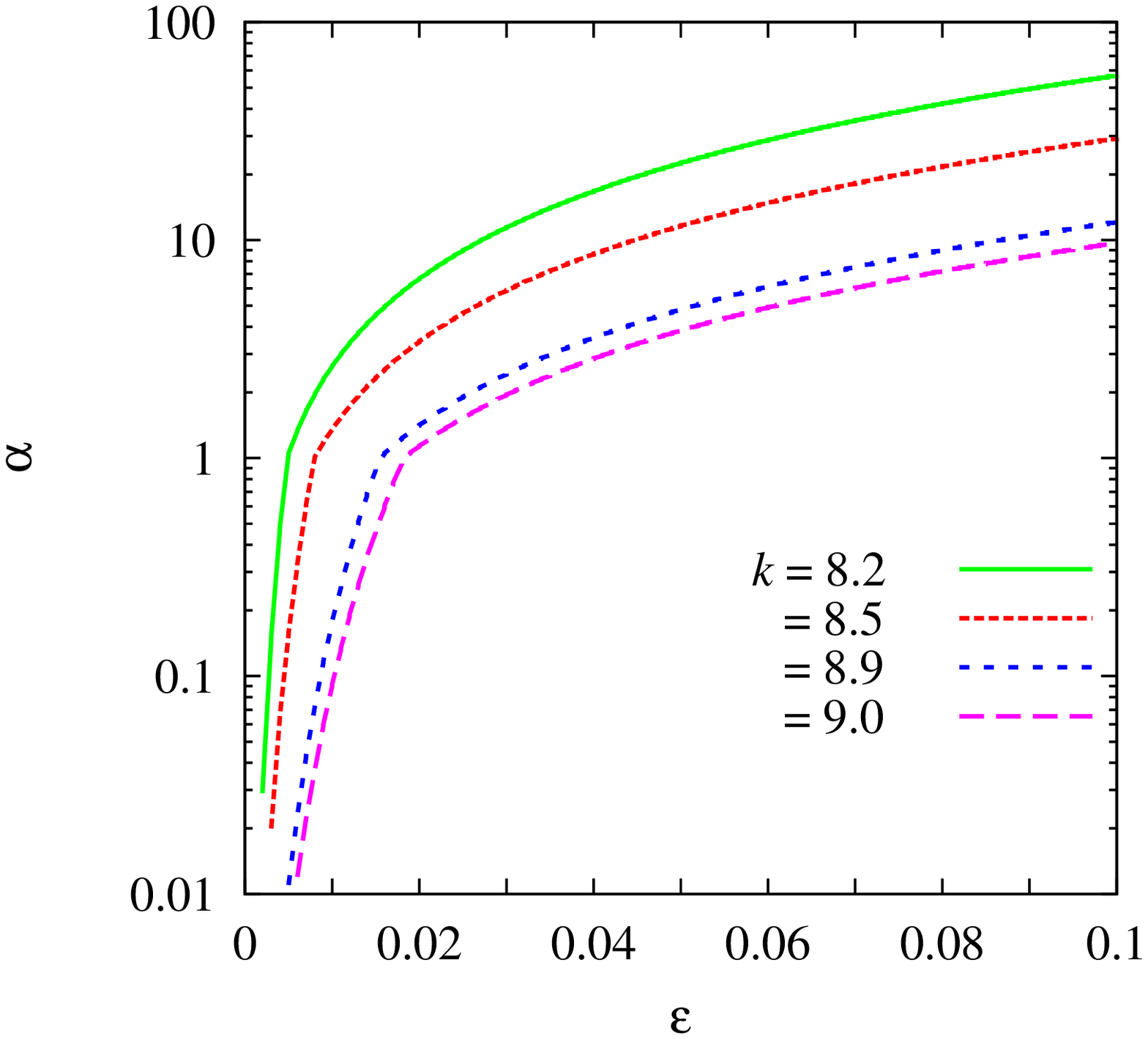}
\epsfxsize=6cm\epsfbox{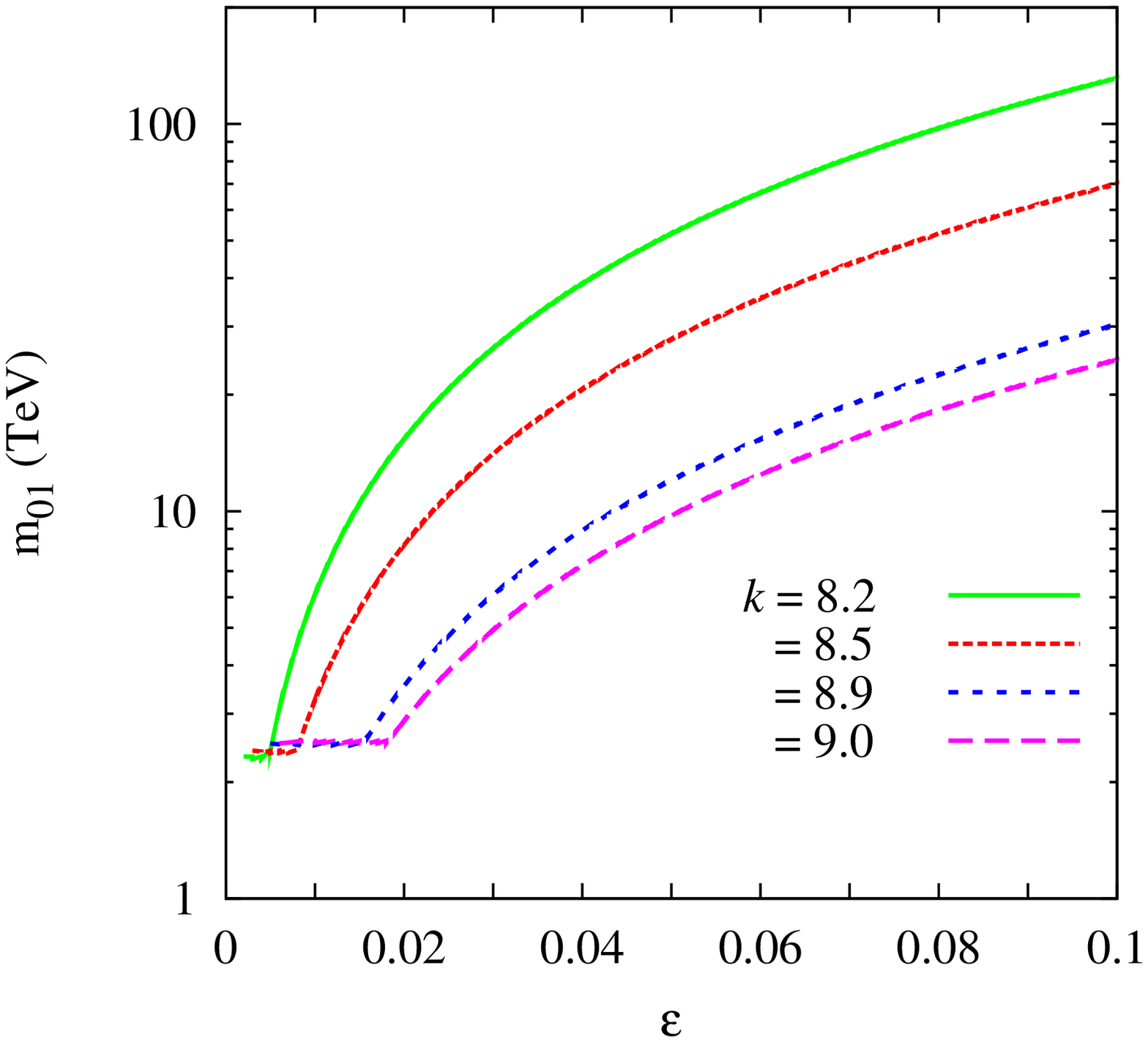}

\vspace*{-60pt}
}
\caption{Left panel: Contour plots in the $(\epsilon, \alpha)$ plane
for fixed values of $k$.  Right panel: The mass $m_{01}$ for the first
graviton mode as a function of $\epsilon$ for a fixed $k$.  The curves
are constrained to satisfy $\mbox{min}(R_y^{-1}, r_z^{-1}) = m_h
/ w$}.
\label{fig:large_k_mass&coup}
\end{figure}

In Fig.\ref{fig:large_k_mass&coup}, we present the interrelationship
between the couplings for various choices of $k$. As in small $k$
sector, here too $\alpha$ increases (decreases) monotonically with $\epsilon$ 
($k$) with the $k$-dependence being much stronger.  
As was expected from the Tables, the typical values of the modulus ratio
$\alpha$ tends to be smaller for this sector. The bend in the curves
(see the left panel) at $\alpha = 1$ are a consequence of our
assertion that the cutoff of the four-dimensional theory is given by
$\mbox{min}(R_y^{-1}, r_z^{-1})$, thereby changing the parametric
dependence of the hierarchy at this point\footnote{Note that $\alpha <
1$ was impossible to obtain in the small $k$ sector.}. Naturally, this
change is also manifested in the relation between $\alpha$ and
$\epsilon$ in the shape of very sharp bends (with the position of the
bend being given by $\alpha(\epsilon, k) = 1$). Below this point, the
mass of the first KK mode, $h^{(0,1)}$ in the case, is almost
independent of $\epsilon$, and is given essentially in terms of
$R_y^{-1}$, which, of course, is determined once the Higgs mass and
the hierarchy determinator $k$ are fixed. A further feature of this
sector is that the coupling $C_{01}$ is essentially fixed by $k$ alone
with only a very subdominant dependence on $\epsilon$.

\section{Discussion and Summary}

Within the original (five-dimensional) RS scenario, the masses and
couplings of the graviton KK modes are determined in terms of very few
tunable parameters. Exploiting this, the ATLAS group searched for the
existence of a graviton resonance in the dilepton mode, and has ruled
out the existence of any such mode below $\sim 2.2$ TeV as long as it
couples to the SM fields with a strength of the order of an inverse
TeV. This negative result is in direct conflict with the RS mechanism's
resolution of the mass hierarchy problem. Thus, one is forced to
accept at least a partial hierarchy, whether it be applicable to the
low energy theory or whether it appears in the guise of an ad hoc
introduction of a scale (for the four-dimensional theory) at
least two orders lower than the natural scale of the problem.

Since neither of these solutions are particularly attractive given the
great promise of the RS paradigm, we have striven here to offer an
alternative and natural solution. The key is the generalization to 
dimensions larger than five and admit multiple warping. Such a situation,
of course, is not unexpected within, say, a string theoretic framework.

While the number of extra dimensions (and independent warpings) can be
arbitrary~\cite{dcssg}, we have restricted ourselves, for reasons of
simplicity, to the six-dimensional theory with two subsequent warpings
and orbifoldings.  This immediately introduces some extra tunable
parameters in the shape of moduli and/or extent of warping.  Further
generalization is straightforward and only serves to increase the
parameters. It should be noted at this stage that the reconciliation
of the ATLAS bounds with the resolution of the hierarchy problem does
not need any extreme tuning of these parameters. Rather, the natural
values of the parameters serve to resolve the conflict.

In a multiple moduli warped model, such as the one under discussion,
it would be advisable to restrict the hierarchy between them to as
small a value as possible. This is over and above maintaining the
smallest of them to be close to the fundamental length scale of the
problem. This serves to maximize the stability of the ratios against
radiative corrections, or, in other words, prevents the reappearance
of the hierarchy problem in a different guise. Such a requirement
forces us to have large warping in only one direction. In other words,
we can have either a large $c$ ($\sim 10$) and a small $k$ ($\lapp
0.3$) or large $k$ ($\gapp 8$) and an almost infinitesimally small
$c$.

The first scenario (large $c$) requires a moderately large ($\gapp
40$) hierarchy between the moduli. This small hierarchy is minimized
by assuming the largest possible ratio between the bulk curvature and
the fundamental mass scale (i.e., the largest $k$).  While, at first
sight, this scenario might seem to be a small perturbation of the
5-dimensional RS model, it is not really so. For one, the graviton
masses are typically larger than those in the RS model, and,
simultaneously have much smaller couplings. Thus, it is almost
straightforward to evade the ATLAS bounds. However, the next run of
the LHC should be able to find them. Even more interestingly, we now
have a double tower of gravitons. In other words, there is a cluster
of relatively closely placed resonances, each with enhanced (to at
least the same level as the first KK mode) couplings waiting to be
discovered at the forthcoming runs of the LHC. And, increasing the
number of warped directions would only serve to increase the density
of these excitations, thereby making the situation quite
lively. Indeed, if we admit as many as 6 extra dimensions, it is
conceivable that these modes can, in the collider environment, start
to mimic a pseudo-continuum of resonances.

The second branch (large $k$) is potentially even more interesting.
For one, it can admit essentially no hierarchy between the moduli.
Essentially only one tower is germane to low energy physics, and the
spacing between the levels is minimized by minimizing the moduli
hierarchy. Even though the modes tend to be somewhat heavier than
those in the RS (thereby largely escaping the ATLAS bounds), the
couplings are no longer suppressed. Thus, a reanalysis of even the
present data can serve to rule out part of the parameter space.

This branch, thus, seems to be an even smaller perturbation of the 
RS, or more correctly, a marriage of the RS with a very small 
ADD-like direction. However, the extremely tiny warping has a profound 
role to play. For one, it is this that allows the 4-brane at $z = 0$ 
(on which our 3-brane is located) to be tensionless. (Compare this 
to the negative tension that the visible brane must have in the 
RS model.) At a phenomenological level, this also serves to bring down 
the fundamental (six-dimensional) mass scale to the GUT scale or 
even below. This is likely to have profound implications 
for model building. Indeed, if we aim to push the fundamental scale close 
to the Planck scale, we enter a strongly coupled phase of the theory! This 
feature is a stark departure from the usual RS scenario.

In summary, we have shown that augmenting the RS scenario 
by incorporating even a single slightly warped extra dimension 
can lead to profound implications. Not only are the current collider 
bounds avoided (though, with the promise of very interesting physics
in the next run of the LHC), but a host of new and exciting 
features emerge.

\section*{Acknowledgement}
   MTA would like to thank UGC-CSIR, India for assistance under Senior Research Fellowship Grant   
     Sch/SRF/AA/139/F-123/2011-12

\end{document}

%% file: Tab_small_k.tex
\begin{table}[!h]
\begin{tabular}{ccc}
$
\begin{array}{|c|c|r|}
\multicolumn{3}{c}{\underline{k = 0.05, \; \alpha  = 211, \; 
       w = 6.14 \times 10^{-15}}}
\\[1ex]
\hline
(n,p) & m_{np} (\tev) & C_{np} \times 10^3 \\ 
& & (\tev^{-1}) \\
\hline 
(1,0)&  5.07   & 8.04  \\ 
\hline 
(2,0) &  9.29  & 8.04  \\ 
\hline 
(3,0) & 13.5 &  8.04  \\ 
\hline 
(0,1)&  30.2   & -24.1   \\ 
\hline 
(1,1) &  37.1  &    16.4   \\ 
\hline 
(2,1) & 42.7  & -14.7    \\ 
\hline 
\end{array}
$
& \hspace*{3em} &
$
\begin{array}{|c|c|r|}
\multicolumn{3}{c}{\underline{k = 0.1, \; \alpha  = 108, \; w  =  8.75 \times 
10^{-15}}}
\\[1ex]
\hline
(n,p) & m_{np} (\tev) & C_{np} \times 10^3 \\ 
& & (\tev^{-1}) \\
\hline 
(1,0)&  5.20   & 5.44   \\ 
\hline 
(2,0) &  9.53   & -5.44\\ 
\hline 
(3,0) & 13.8  &  5.44 \\ 
\hline 
(0,1)&  17.1   & 13.4  \\ 
\hline 
(1,1) &  23.0  &    -9.99 \\ 
\hline 
(2,1) & 28.1  & 9.20  \\ 
\hline 
\end{array}
$
\\[1ex]
& & \\[1ex]
$
\begin{array}{|c|c|r|}
\multicolumn{3}{c}{\underline{k = 0.2, \; \alpha  = 60.9, \; w  =  1.31 \times 
10^{-14}}}
\\[1ex]
\hline
(n,p) & m_{np} (\tev) & C_{np} \times 10^3 \\ 
& & (\tev^{-1}) \\
\hline 
(1,0)&  5.87   & 3.16   \\ 
\hline 
(2,0) &  10.7   & -3.16 \\ 
\hline 
(3,0) & 15.6  &  3.16\\ 
\hline 
(0,1)&  11.6   & 7.19  \\ 
\hline 
(1,1) &  17.4  & -5.93\\ 
\hline 
(2,1) & 22.7  &  5.64 \\ 
\hline 
\end{array}
$

&  &
$
\begin{array}{|c|c|r|}
\multicolumn{3}{c}{\underline{k = 0.3, \; \alpha  = 49.3, \; w  =  1.81 \times 
10^{-14}}}
\\[1ex]
\hline
(n,p) & m_{np} (\tev) & C_{np} \times 10^3 \\ 
& & (\tev^{-1}) \\
\hline 
(1,0)&  7.07   & 1.87 \\ 
\hline 
(2,0) &  12.9  &   -1.87\\ 
\hline 
(3,0) & 18.8 &  1.87\\ 
\hline 
(0,1)&  11.3  & -4.74  \\ 
\hline 
(1,1) &  17.8  & 4.13 \\ 
\hline 
(2,1) & 24.0  &  -3.99 \\ 
\hline 
\end{array}
$

\end{tabular}
\caption{\em Four sample spectra for the small $k$ case 
for a particular bulk curvature ($\epsilon = 0.0775$).}
\label{tab:small_k}
\end{table}

%% file: Tab_large_k_1.tex
\begin{table}[!h]
\begin{tabular}{ccc}
$
\begin{array}{|c|c|r|}
\multicolumn{3}{c}{\underline{k=8.2, \; \alpha = 9.87, \; \epsilon = 0.027}}
\\[1ex]
\hline
\multicolumn{3}{|c|}{w = 1.3 \times 10^{-11} }
\\
\hline
(n,p) & m_{np} (\tev) & C_{np} (\tev^{-1}) \\ 
\hline 
(0,1) &  22.98    &  -0.881 \\ 
\hline 
(0,2) & 47.09    & 0.745 \\ 
\hline  
(0,3) &  68.94  &  -0.720 \\ 
\hline 
(0,4) & 90.17  & 0.710 \\ 
\hline 
\end{array}
$
& \hspace*{3em} &
$
\begin{array}{|c|c|r|}
\multicolumn{3}{c}{\underline{k=8.5, \; \alpha = 9.87, \; \epsilon = 0.044}}
\\[1ex]
\hline
\multicolumn{3}{|c|}{w = 5.06 \times 10^{-12}}
\\
\hline
(n,p) & m_{np} (\tev) & C_{np} (\tev^{-1}) \\ 
\hline 
(0,1) & 23.35   &  -3.62 \\ 
\hline 
(0,2) & 47.86    & 3.06  \\ 
\hline  
(0,3) & 70.07 &  -2.96  \\ 
\hline 
(0,4) & 91.65  & 2.92 \\ 
\hline 
\end{array}
$
\\[2ex]
& & 
\\[2ex]
$
\begin{array}{|c|c|r|}
\multicolumn{3}{c}{\underline{k = 8.2, \; \alpha  = 1.56, \; 
       \epsilon = 0.00675}}
\\[1ex]
\hline
\multicolumn{3}{|c|}{w = 1.3 \times 10^{-11}}
\\
\hline
(n,p) & m_{np} (\tev) & C_{np} (\tev^{-1}) \\ 
\hline 
(0,1)&  3.61   & -0.881   \\ 
\hline 
(0,2) & 7.40   &   0.745  \\ 
\hline 
(0,3) & 10.8 &  -0.720  \\ 
\hline 
(0,4) & 14.2 &  0.710   \\ 
\hline 
\end{array}
$
& & 
$
\begin{array}{|c|c|r|}
\multicolumn{3}{c}{\underline{k = 8.5, \; \alpha  = 1.56, \; \epsilon = 0.0111}}
\\[1ex]
\hline
\multicolumn{3}{|c|}{w = 5.06 \times 10^{-12}}
\\
\hline
(n,p) & m_{np} (\tev) & C_{np} (\tev^{-1}) \\ 
\hline 
(0,1)&  3.74   & -3.62  \\ 
\hline 
(0,2) & 7.66  &  3.06  \\ 
\hline 
(0,3) & 11.2 &  -2.96  \\ 
\hline 
(0,4) & 14.7 &  2.92 \\ 
\hline 
\end{array}
$
\end{tabular}
\caption{\em Four sample spectra for the large $k$ case.}
\label{tab:large_k_1}
\end{table}

%% file: Tab_large_k_2.tex
\begin{table}[!h]
\begin{tabular}{ccc}
$
\begin{array}{|c|r|r|}
\multicolumn{3}{c}{\underline{k =8.9, \; \alpha = 1.56, \; \epsilon =0.021}}
\\[1ex]
\hline
\multicolumn{3}{|c|}{w = 1.44 \times 10^{-12}}
\\
\hline
(n,p) & m_{np} (\tev) & C_{np} (\tev^{-1}) \\ 
\hline  
(0,1) &  3.87   &  -23.9 \\ 
\hline 
(0,2) & 7.92  & 20.2  \\ 
\hline  
(0,3) & 11.59 &  -19.5   \\ 
\hline 
(0,4) & 15.16  & 19.2 \\ 
\hline 
\end{array}
$
& \hspace*{3em} &
$
\begin{array}{|c|r|r|}
\multicolumn{3}{c}{\underline{k =11, \; \alpha = 0.002, \; \epsilon =0.1}}
\\[1ex]
\hline
\multicolumn{3}{|c|}{w = 1.96 \times 10^{-15}}
\\
\hline
(n,p) & m_{np} (\tev) & C_{np} (\tev^{-1}) \\ 
\hline 
(0,1) &  3.20   &   -4.29 \times 10^5\\ 
\hline 
(0,2) &  6.56  & 3.62 \times 10^5\\ 
\hline 
(0,3) &  9.59  &  -3.50 \times 10^5\\ 
\hline 
(0,4) &  12.54  &  3.45 \times 10^5\\ 
\hline
\end{array}
$
\end{tabular}
\caption{\em Two additional sample spectra for the large $k$ case.}
\label{tab:large_k_2}
\end{table}